\documentclass[aps, llncs, preprint]{revtex4-1}
\usepackage{graphicx}
\usepackage{amsmath}
\usepackage{float}
\usepackage{color}
\usepackage[dvipsnames]{xcolor}
\usepackage{tabularx}
\usepackage{subfiles}

\begin{document}

\title{Topological surface states host superconductivity induced by the bulk condensate in YRuB$_2$}

\author{Nikhlesh Singh Mehta$^1$}
\author{ Bikash Patra$^2$, Mona Garg$^1$, Ghulam Mohmad$^1$, Mohd Monish$^1$, Pooja Bhardwaj$^1$, P. K. Meena$^3$, K. Motla$^3$, Ravi Prakash Singh$^3$, Bahadur Singh$^2$}
\author{Goutam Sheet$^1$}
\email{goutam@iisermohali.ac.in}
\affiliation{$^1$Department of Physical Sciences, Indian Institute of Science Education and Research (IISER) Mohali, Sector 81, S. A. S. Nagar, Manauli, PO 140306, India}
\affiliation{$^2$Department of Condensed Matter Physics and Materials Science, Tata Institute of Fundamental Research, Mumbai 400005, India}
\affiliation{$^3$Department of Physics, Indian Institute of Science Education and Research Bhopal, Bhopal 462066, India}

\begin{abstract}
   
While the possibility of topological superconductivity (TSC) in hybrid heterostructures involving topologically nontrivial band structure and superconductors has been proposed, the realization of TSC in a single stoichiometric material is most desired for fundamental experimental investigation of TSC and its device applications. Bulk measurements on YRuB$_2$ detect a single superconducting gap of $\sim$ 1 meV. This is supported by our electronic structure calculations which also reveal the existence of topological surface states in the system. We performed surface-sensitive Andreev reflection spectroscopy on YRuB$_2$ and detected the bulk superconducting gap as well as another superconducting gap of $\sim$ 0.5 meV. From our analysis of electronic structure, we show that the smaller gap is formed in the topological surface states in YRuB$_2$ due to the proximity of the bulk superconducting condensate. Thus, in agreement with the past theoretical predictions, we present YRuB$_2$ as a unique system that hosts superconducting topological surface states.
    
\end{abstract}
\maketitle

Due to the particle-hole symmetry in superconductors, the positive and negative energy eigenstates of the Bogoliubov-DeGennes Hamiltonian appear pair-wise \cite{Alicea}. When the superconducting condensate forms, the negative-energy eigenstates remain fully occupied. This closely resembles to the insulators where the valence band remains filled. Therefore, distinct topological invariants for the occupied states can be calculated \cite{kane and Mele, Fu and kane,
Hasan and kane, Moore}. A non-zero topological invariant leads to a so-called
topological superconductor. In strong topological superconductors, the non-zero topological invariants may exist even when the bulk of the superconductor is fully-gapped, conventional\cite{Sato, Qi,Schnyder}. Due to the constraints enforced by topology, the surface of the strong topological superconductors host gap-less modes, the so-called Majorana zero modes. Such systems have recently attracted enormous attention due to their fascinating properties and their potential as a key ingredient of fault-tolerant quantum computing \cite{Nayak, sharma, sharma2}. Therefore, it is extremely important to search for candidate topological superconductors (TSCs).

Fu and Kane had proposed a scheme to obtain $p_x + ip_y$ type topological superconductivity induced in the topological surface states (TSSs) of a topological insulator (TI) through proximity effect, or by doping \cite{Fu and kane2}. Experimentally, proximity induced superconductivity was observed in heterostructures of superconducting NbSe$_2$, and BSCCO with the 
topological insulator Bi$_2$Se$_3$ \cite{wang,wang2}. Bi$_2$Se$_3$ was intercalated with metal ions like Cu\cite{wray}, Nb\cite{Asaba}, Sr \cite{Liu} etc. in a controlled way to achieve superconductivity. Similarly, upon In doping, the topological crystalline insulator SnTe displayed superconductivity \cite{satoshi}. In all such cases, the intrinsic features of a TSC may undergo modification due to complex interface effects, strains developing from lattice mismatched heterostructures or the intercalates acting as disorder. All such issues can be overcome only if a TSC phase is realized in a single stoichiometric material system. In such a system, the Majorana zero modes may appear as exotic surface states, or as bound states in the vortex cores. As per the theoretical argument presented in \cite{Gao}, a TSS with a reasonably high $T_c$ is required to experimentally resolve the Majorana bound states. However, till date, all the stoichiometric topological systems that have shown superconductivity have rather low $T_c$ (e.g., the Dirac semimetal PdTe$_2$ ($T_c=$ 1.7 K)\cite{Noh}, the nodal line semimetal PbTaSe$_2$ ($T_c=$ 3.8 K)\cite{chang}, BiPd ($T_c=$ 3.8 K))\cite{sun,Mondal}. \par
Recently, based on electronic structure calculations\cite{Gao}, it was proposed that the rare-earth transition-metal
ternary boride YRuB$_2$ ($T_c =$ 7.6 K) is a ideal topological superconductor candidate where all the key requirements for being a topological superconductor (namely, a relatively higher $T_c$, topological surface states, $s$-wave superconductivity and good separation between bulk and surface states) are satisfied. In addition to the above, the calculations also suggest the existence of symmetry-protected Dirac nodal rings in YRuB$_2$\cite{STM}. Motivated by such theoretical observations, we have employed Andreev reflection spectroscopy \cite{Andreev,Naidyuk} experiments on YRuB$_2$. Andreev reflection spectroscopy is known to be a potentially powerful technique to probe transport through topological surface states in a topological superconductor\cite{sasaki, sun2, Dai}.

\begin{figure}[htp]
 \includegraphics[scale = 0.45]{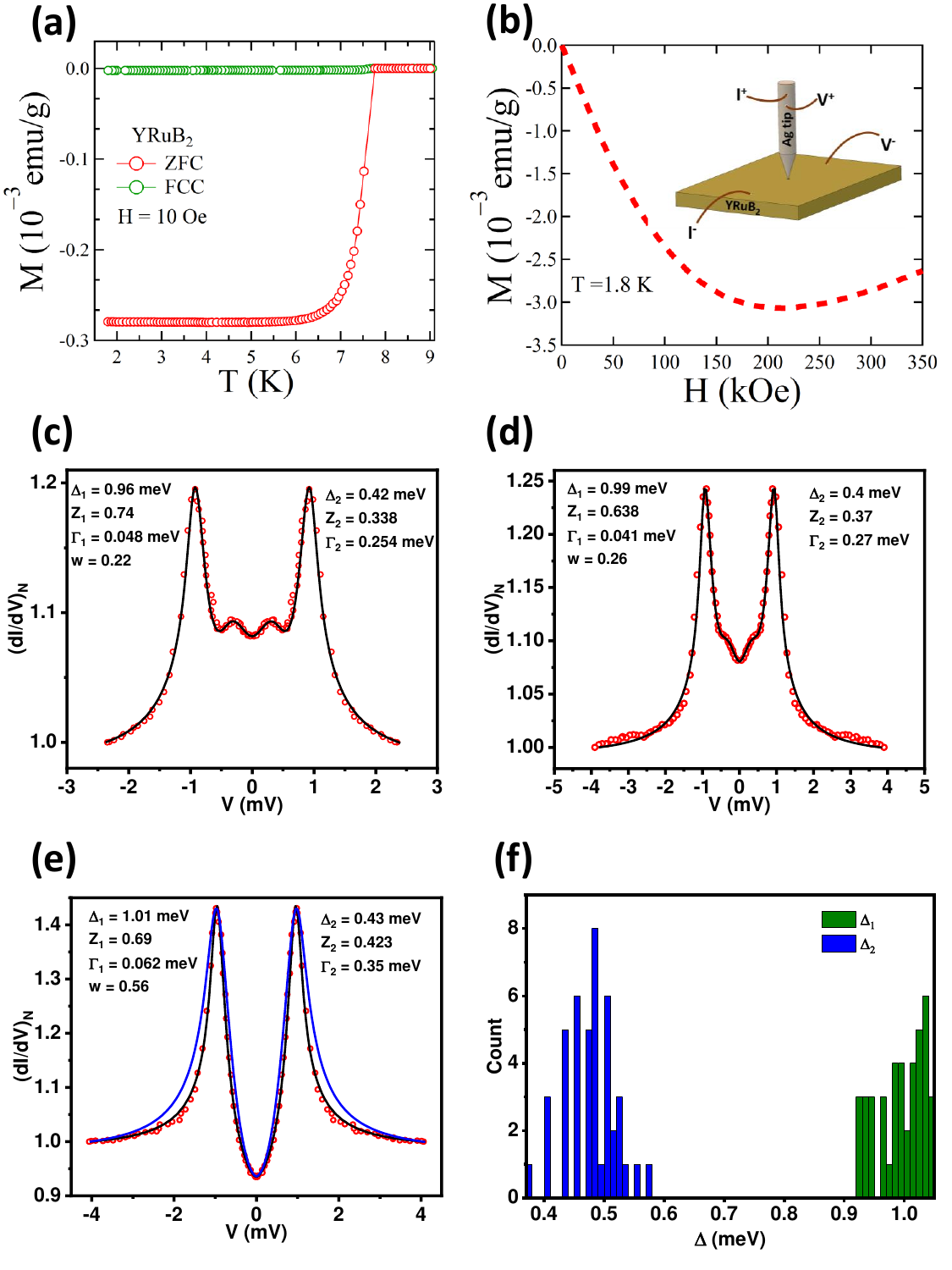}
 \caption{(a) Temperature dependence of bulk magnetization measured in both zero field cooling (ZFC) and field cooled cooling mode with 10 Oe magnetic field (FCC). (b) Field dependence of bulk magnetization at a temperature of 1.8 K.  Inset shows a schematic describing the formation of point contact on YRuB$_2$. (c,d) Conductance spectra obtained in the ballistic regime (red circles) and their corresponding two-gap BTK fit (black line). (e) Conductance spectra obtained in the ballistic regime (red circles) and their corresponding single gap BTK fit (blue line) and two gap fit (black line). (f)  Statistics of superconducting energy gaps ($\Delta_1$ and $\Delta_2$ ) for 43 different contacts.
  }
  \label{fig_1}
\end{figure}

The measurements reported here were performed on polycrystalline YRuB$_2$ where multiple single crystallites with randomly oriented facets coexist on the surface. Temperature dependence of bulk magnetization on polycrystalline YRuB$_2$ in zero field cooled (ZFC) and field cooled cooling (FCC) mode with an applied field of 10 Oe confirm the bulk nature of superconductivity in YRuB$_2$ and it shows a superconducting transition onset at 7.8 K as shown in Figure 1(a). The Andreev reflection spectroscopic measurements were performed by measuring the transport characteristics of several ballistic point-contacts between superconducting YRuB$_2$ and normal metallic Ag tips respectively. 
\par

The electronic transport between a normal metal and a superconductor through a ballistic point contact is dominated by a process called Andreev reflection\cite{Andreev} that causes an enhancement of the differential conductance ($dI/dV$) when the electron energy is less than the superconducting energy gap ($\Delta$). A $dI/dV$ vs energy ($E = eV$) spectrum thus obtained is analyzed by a modified Blonder-Tinkham- Klapwijk (BTK) model\cite{BTK}. This model assumes the interface between a normal metal and a superconductor as a delta potential barrier whose strength is defined by a dimensionless parameter $Z$. With a small potential barrier present at the interface, two peaks symmetric about $V$=0 appear. Such peaks are the hallmark signatures of Andreev Reflection. In Figure 1(c-e), three representative point contact Andreev reflection (PCAR) spectra between YRuB$_2$ and Ag probed at T $\sim$ 0.45 K are shown (red circles). All the spectra were first normalized with respect to the conductance at high bias. In all these spectra, Andreev peaks symmetric about $V$=0, are clearly seen. No extra (anomalous) features like conductance dips\cite{Goutam,kumar} are present. This confirms that the point contacts are close to the ballistic regime of transport where true spectroscopic parameters can be obtained. We have performed such experiments at a large number of points (see Figure S9 - S12 in supplementary information). The normal state resistance of these points varied from 0.6 $\Omega$ to 20 $\Omega$ and the contact diameters (calculated using Wexler's formula\cite{Wexler}) varied between 4 nm and 24 nm. As shown in Figure 1(c,d), there are two well resolved peaks for both positive and negative $V$ in the point contact spectra obtained on YRuB$_2$. This is strikingly similar to the Andreev reflection spectra obtained on the two-band superconductor MgB$_2$\cite{Gonnelli,Szabo}. The solid black lines in Figure 1(c,d) represent the theoretical fits using the modified BTK theory generalized to include two superconducting gaps ($\Delta_1$ and $\Delta_2$) by writing the normalized conductance as $(\frac{dI}{dV} )_N$ = $w(\frac{dI}{dV} )_{1N} + (1-w)$  $(\frac{dI}{dV} )_{2N}$, where $w$ is the relative contribution of one of the gaps (say, $\Delta_1$)\cite{Gonnelli}.  
As shown in Figure 1(e), for certain point contacts, we also obtained spectra in which two gaps are not visually resolved. We noted that while a conventional single-gap model is insufficient to explain these spectra, the two-gap model provides a better fit to the spectra over the entire energy range. We thus obtained a distribution of the two gaps measured at different points on the surface of YRuB$_2$ and plotted the distribution in Figure 1(f). As it is evident from the distribution, prominently two superconducting gaps are measured with $\Delta_1 =$ 0.99 $\pm 
 $ 0.07 meV and $\Delta_2 =$ 0.47 $\pm$ 0.1 meV.
\begin{figure}[h!]
\centering
\includegraphics[width=0.47\textwidth]{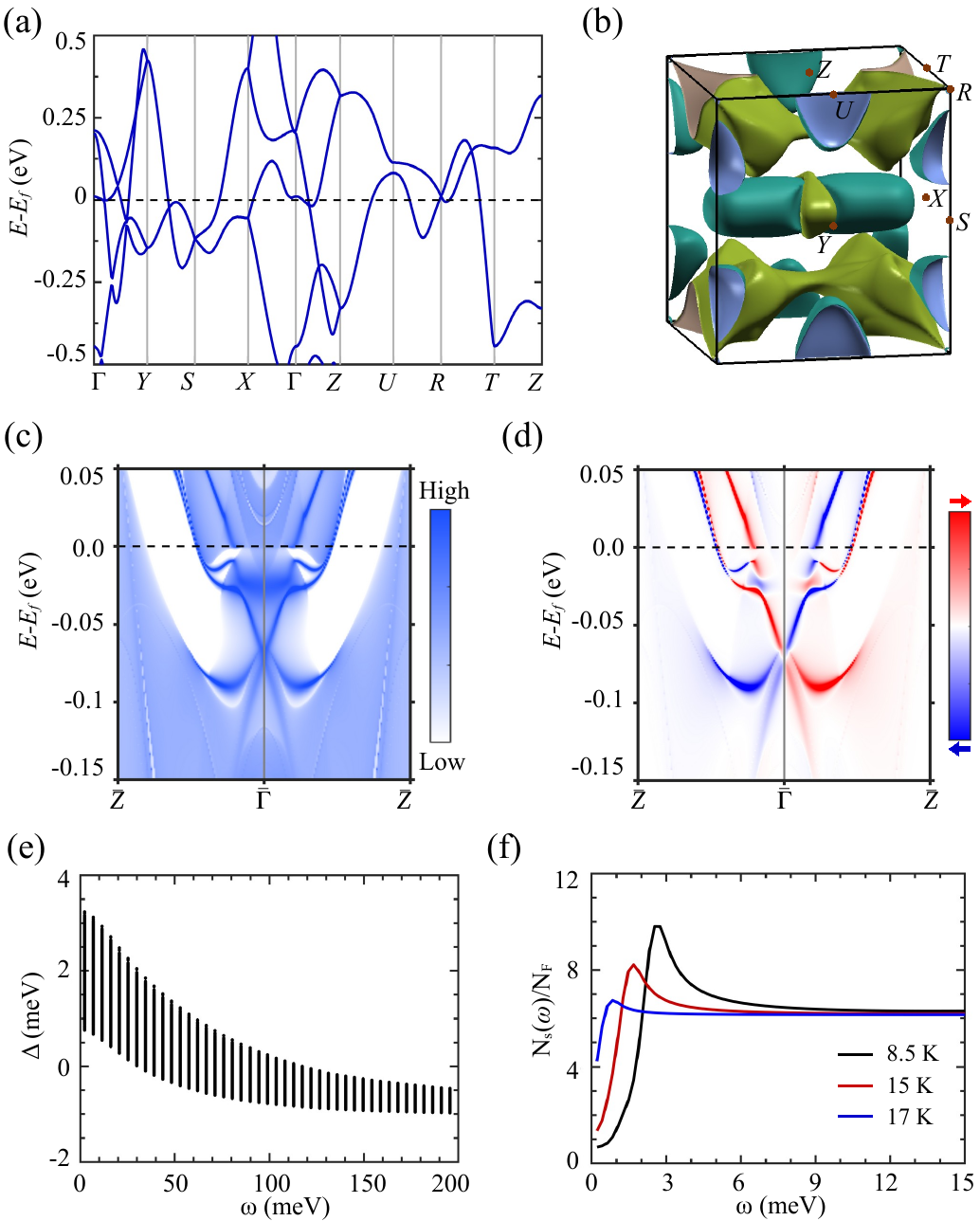}
\caption{(a) Calculated bulk band structure of YRuB$_2$ without SOC along the high-symmetry path in the Brillouin zone (BZ). (b) The associated Fermi surface with coexisting electrons (green and brown colors) and hole pockets (light blue and teal colors). (c) The (010) surface band structure along $\bar{Z}-\bar{\Gamma}-\bar{Z}$ directions. (d) The associated spin texture with up (red) and down (blue) spin polarizations. (e) Superconducting gap along the imaginary axis at $T=8.5$ K and (f) the quasiparticle superconducting density of states (DOS) at $8.5$ K, $15$ K, and $17$ K.}
\label{fig_bands}
\end{figure}
 
The above observation substantially differs from the earlier measurements of the superconducting energy gap in YRuB$_2$ based on bulk measurement techniques like NMR relaxation and $\mu$-SR experiments in the past\cite{NMR,Barker}. Both these experiments revealed one clean, fully formed superconducting gap in YRuB$_2$ with an amplitude of $\sim$ 1.1 meV that followed BCS behaviour\cite{BCS}. In our experiments, the measured larger gap ($\Delta_1$) is comparable to the bulk gap amplitude reported by the other bulk-sensitive experiments\cite{NMR,Barker}. To understand the origin of the smaller gap ($\Delta_2$) in our measurements, we have performed detailed first-principles calculations.
We presented the calculated bulk band structure of YRuB$_2$ without spin-orbit coupling (SOC) in Figure~\ref{fig_bands}(a). It manifests a metallic ground state where electron and hole bands dip into each other in such a way that they form coexisting electron and hole pockets at the Fermi level. Figure~\ref{fig_bands}(b) shows the calculated Fermi surface that reveals two hole pockets (light blue and teal colors) and two electron pockets (green and brown colors), where a single superconducting gap forms (please 
 also see supplementary information for additional details). Since YRuB$_2$ has non-symmorphic symmetries, its metallic state is robust and realizes hourglass Dirac Fermions at the zone boundary in the presence of SOC. Regardless, there is a band inversion between the valence and conduction bands at the $\Gamma$ point such that the valence and conduction bands are separated at $k_y = 0$ plane. Such a gapped state can facilitate the calculations of $Z_2$ number on these planes similar to insulators. Based on the parity eigenvalues of the occupied states, we obtained a nontrivial $Z_2=1$ on the  $k_y=0$ plane. Figure~\ref{fig_bands}(c) shows the calculated (010) surface states and associated spin-texture in Figure ~\ref{fig_bands}(d). These results reveal an odd number of spin-momentum locked nontrivial states crossings along $\bar{\Gamma}-\bar{Z}$ at the Fermi level. Such nontrivial states can in principle become superconducting through the bulk proximity effect.
Since point contact spectroscopy is a more surface sensitive technique, the Andreev reflection processes in our experiments are bound to involve the bulk gap as well as the proximity induced gap in the TSSs. As a consequence, we have effectively measured two gaps in YRuB$_2$. We also computed the phonon dispersion and Eliashberg spectral function $\alpha^2F(\omega)$ to get the superconducting $T_c$ (see Figure S2 in supplementary information).
The calculated value of $T_c$ using the McMillan formula as modified by Allen and Dynes~\cite{Allen} is $8.7$ K. 
Figure~\ref{fig_bands}(e) shows the superconducting gap function at 8.5 K  obtained by solving the (anisotropic) Eliashberg equation along the imaginary axis. Quasiparticle density of states $N_s(\omega)/N(E_F)={\rm Re}[\omega/\sqrt{\omega^2-\Delta^2(\omega)}]$, where $N(E_F)$ is the normal density of states (DOS) at the Fermi level, in the superconducting state is shown in Figure~\ref{fig_bands}(f). A single peak in the quasiparticle DOS signifies the presence of only a single bulk superconducting gap. The peak in the DOS gradually disappeared above 17 K. This overestimated temperature scale might be due to the possible anharmonic effects~\cite{floris2005superconducting}, or the use of an isotropic Coulomb parameter~\cite{choi2002first}. Nevertheless, considering the possibility of only one bulk superconducting gap, it is rational to surmise that the second gap measured by our experiments is a proximity-induced gap in the surface states. 

\begin{figure}
 \centering
 \includegraphics[width= 0.45\textwidth]{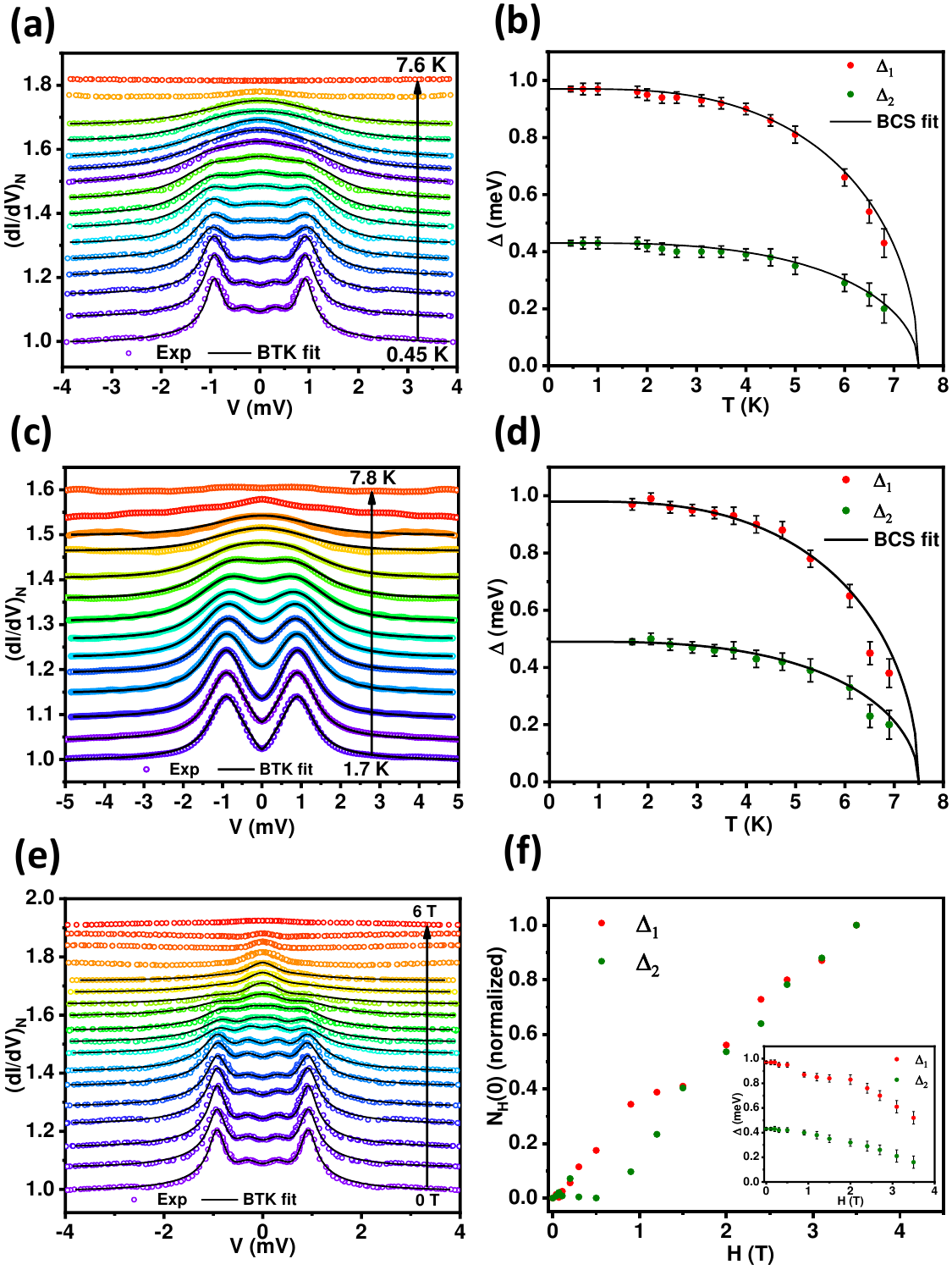}
 \caption{(a) Temperature ($T$) dependence of the conductance spectra (colored circles) with two gap BTK fit (black line). (b) Variation of the two superconducting gaps ($\Delta_1$ and $\Delta_2$ ) with temperature ($T$). (c) $T$ dependence of the conductance spectra of other type where two gaps are not visually resolved. (d) Evolution of  $\Delta_1$ and $\Delta_2$, extracted from the spectra shown in Figure 3(c), with $T$. (e) Magnetic field ($H$) dependence of the conductance spectra (colored circles) with two gap BTK fit (black line) at 0.45 K. (f) $H$ dependence  of zero bias density of states ($N_H(0)$) after subtracting the zero field contribution corresponding to $\Delta_1$ and $\Delta_2$. The $inset$ shows the variation of the two superconducting gaps ($\Delta_1$ and $\Delta_2$ ) with $H$.
  }
  \label{fig_3}
\end{figure}

Now it is important to investigate the nature of the two superconducting gaps. For that, we have investigated the response of the PCAR spectra and the corresponding $\Delta_1$ and $\Delta_2$ with changing temperature and magnetic fields. Figure ~\ref{fig_3}(a) depicts the temperature dependence of $dI/dV$ vs $V$ spectra. The colored circles represent the experimentally registered data points and the solid black lines represent the BTK fits generalized to the case of two gaps for each spectrum. For two gap fitting, the value of the weight factor $w$ was kept fixed over the entire temperature range.  At low temperatures, the position of the Andreev reflection driven peaks does not show a noticeable change. With increasing temperature, the peaks slowly broaden and eventually all spectral features disappear at a temperature of 7.6 K, near the critical temperature of the superconductor. Figure ~\ref{fig_3}(b) shows the temperature dependence of $\Delta_1$ and $\Delta_2$ extracted from the spectrum shown in Figure ~\ref{fig_3}(a). Here, the red and green dots represent the extracted values of $\Delta_1$ and $\Delta_2$ respectively and the solid black lines represent the expected temperature dependence of $\Delta_1$ and $\Delta_2$ for a conventional BCS superconductor\cite{BCS}. Good quality BTK fits of the experimental data and a near-BCS temperature dependence of both the gaps show that the corresponding order parameters are conventional in nature. The extracted value of $\Delta_{01}$ and $\Delta_{02}$ for this particular spectrum are 0.97 meV and 0.43 meV respectively. The corresponding $\frac{2\Delta_0}{K_BT_c}$ for $\Delta_1$ and $\Delta_2$ were found to be $\sim$ 2.96 and 1.32 respectively. The ratio corresponding to the larger gap falls within the weak-coupling BCS regime and is consistent with the previous bulk-sensitive experiments\cite{NMR,Barker}. The temperature dependence of a representative spectrum of the other type where the two gaps are not visually resolved, along with the respective theoretical fits within the two-gap model, is also shown in Figure ~\ref{fig_3}(c). The two gaps extracted from this point also follows the BCS temperature dependence as shown in Figure ~\ref{fig_3}(d).

In order to obtain additional understanding on the two gaps of YRuB$_2$, we performed magnetic field dependence of the PCAR spectra. Figure ~\ref{fig_3}(e) shows the magnetic field dependence of $dI/dV$ vs $V$ spectra. The colored circles represent the experimentally obtained spectra and the solid black lines represent the two gap BTK fits for each spectrum. With increasing magnetic field, Andreev reflection driven conductance peaks close smoothly and all the superconductivity-related features disappear at a magnetic field of $\sim$ 6 T. The extracted value of $\Delta_1$ and $\Delta_2$ for the spectra at zero magnetic field and at a temperature of 0.45 K are 0.97 meV and 0.43 meV respectively. The variation of $\Delta_1$ and $\Delta_2$ with the magnetic field are shown in inset of Figure ~\ref{fig_3}(f). For lower magnetic fields (upto 2 T) $\Delta_1$ and $\Delta_2$ do not change significantly and they decrease smoothly with further increasing magnetic field. Beyond this field, $\Delta_1$ and $\Delta_2$ fall rapidly and a linear extrapolation of data shows that $\Delta_2$ has a tendency to disappear at a magnetic field of $\sim$ 4.8 T, while $\Delta_1$ at $\sim$ 5.8 T.  \par
In the context of the conventional multiband superconductor MgB$_2$, it was earlier shown that the zero bias density of states (DOS) corresponding to the smaller gap grows far more rapidly and attains the normal state value much before that corresponding to the larger gap\cite{Koshelev}. We have calculated the zero-bias density of states (DOS) using the Dyne's formula given as $N(E)=Re\left[\frac{E-i\Gamma}{\sqrt{(E-i\Gamma)^2-\Delta^2}}\right]$\cite{dynes}. 
Magnetic filed dependence of $N(0)$ corresponding to $\Delta_1$ and $\Delta_2$ for the spectra shown in Figure ~\ref{fig_3}(e) is shown in Figure S13 in  supplementary information.
Figure ~\ref{fig_3}(f) shows the corresponding field dependence of $N_H(0)= N(0)-N_0(0)$ where $N_0$ is the zero-bias DOS for $H$ = 0. For YRuB$_2$, it appears that the effect of magnetic field on the DOS corresponding to both the larger and the smaller gap is the same and they evolve with field following a similar trend. This suggests that the two gaps do not independently form in two different bands, but are closely related where the smaller gap is induced in the TSS by the larger one in the bulk. Within this picture, since the amplitude of the proximity-induced gap varies between the crystallite facets, the relatively large distribution of the measured superconducting energy gaps is understood as a consequence of the randomly oriented crystallite facets on the surface of our sample on which the Ag tip falls. 

In conclusion, we have performed point contact Andreev reflection spectroscopy experiments on the candidate topological superconductor YRuB$_2$. While based on bulk measurements YRuB$_2$ is thought to be a single gap superconductor, in our experiments we detected multiple superconducting gaps centered around two amplitudes, 0.99 meV and 0.47 meV. We have shown through the first-principles calculations that the emergence of the smaller gap in our surface sensitive experiments is a consequence of a proximity induced superconducting gap in the TSSs in the system, some of which cross the Fermi surface and contribute in global transport. The properties of the larger gap is consistent with that probed by bulk-sensitive experiments. Therefore, our experiments show that YRuB$_2$ is a potentially important superconductor where the interaction between topological surface states and bulk superconductivity leads to novel physical insights in understanding the candidate topological superconductors.\par
We would like to thanks Ms. Savita and Dr.Yogesh for their help in resistivity measurement. We acknowledge SEM central facility at IISER Mohali. N.S.M. thanks University Grants Commission (UGC) for senior research fellowship (SRF). G.M. thanks UGC for junior research fellowship (JRF). M.G. thanks the Council of Scientific and Industrial Research (CSIR), Government of India, for financial support through a research fellowship (Award No. 09/947(0227)/2019-EMR-I).  M.M. thanks the Council of Scientific and Industrial Research (CSIR), Government of India, for financial support through a research fellowship (Award No. 09/0947(12989)/2021-EMR-I). We thank Sougata Mardanya for fruitful discussions in calculating the superconducting properties. The work at TIFR Mumbai was supported by the Department of Atomic Energy of the Government of India under Project No. 12-R$\&$D-TFR-5.10-0100 and benefited from the computational resources of TIFR Mumbai. R.P.S. acknowledge the Science and Engineering Research Board, Government of India, for the Core Research Grant CRG/2019/001028. G.S. acknowledges financial assistance from the Science and Engineering Research Board (SERB), Govt. of India (grant number: CRG/2021/006395).

\pagebreak

\title{Supplementary information \\
\Large Topological surface states host superconductivity induced by the bulk condensate in YRuB$_2$}

\author{Nikhlesh Singh Mehta$^1$}
\author{ Bikash Patra$^2$, Mona Garg$^1$, Ghulam Mohmad$^1$, Mohd Monish$^1$, Pooja Bhardwaj$^1$, P. K. Meena$^3$, K. Motla$^3$, Ravi Prakash Singh$^3$, Bahadur Singh$^2$}
\author{Goutam Sheet$^1$}
\email{goutam@iisermohali.ac.in}
\affiliation{$^1$Department of Physical Sciences, Indian Institute of Science Education and Research (IISER) Mohali, Sector 81, S. A. S. Nagar, Manauli, PO 140306, India}
\affiliation{$^2$Department of Condensed Matter Physics and Materials Science, Tata Institute of Fundamental Research, Mumbai 400005, India}
\affiliation{$^3$Department of Physics, Indian Institute of Science Education and Research Bhopal, Bhopal 462066, India}
\maketitle
\section{ Methods:}
Polycrystalline YRuB$_2$ samples were synthesized using an arc-melting technique. High-purity Y, Ru, and B elements were combined in stoichiometric ratios and melted in a tetra-arc furnace. The melting process was carried out under an argon (5N) atmosphere on a water-cooled copper hearth. To enhance the homogeneity of the initial casting, the samples were flipped and remelted multiple times. Afterwards, the samples were carefully sealed in evacuated quartz tubes and subjected to annealing at 1050°C for one week.\par
\setlength{\parindent}{10pt}
The Andreev reflection spectroscopic measurements were performed by using a home-built point-contact spectroscopy(PCS) probe down to 450 mK in a He-3 cryostat\cite{Shekhar}. 
The point contacts between YRuB$_2$ and Ag were formed using needle-anvil technique. First, the surface of YRuB$_2$ was polished and mounted on a sample stage in the home-built PCS probe. A calibrated ruthenium oxide thermometer and a heater were attached to the sample stgae for accurate measurement and control of temperature. The sample was placed at the center of a 7 T magnet and all the magnetic field measurements were performed by applying a magnetic field perpendicular to the sample surface. Andreev reflection spectroscopic measurements at 1.8 K were performed using a home-built PCS probe in a liquid helium cryostat.\par
First-principles calculations were performed within the density functional theory framework using the Quantum Espresso package with the norm-conserving pseudopotentials~\cite{QE,NC_TM, NC_pseudojo}. We employed local density approximation (LDA) to incorporate exchange-correlation effects~\cite{LDA_1980, LDA_perdew1981}. We obtained the phonon spectrum using the density-functional perturbation theory. The plane-wave kinetic energy cut-off was used as 60 Ry. The electron-phonon coupling (EPC) calculations were performed to evaluate the Eliashberg spectral function and superconducting $T_c$ using the EPW code with a Wannier interpolated fine 32$\times$32$\times$32 $\bf{k}$-grid and 16$\times$16$\times$16 $\bf{q}$-grid~\cite{EPW1, EPW2}.
\section{Experimental details:}
There are several ways to create a point contact between two different materials. One of the simplest is the needle anvil technique in which a point contact can be formed by simply touching a sharp metallic tip to the sample. In this technique, the movement of the tip can be controlled either using a piezo-controlled nanopositioner or a  mechanical differential screw arrangement. 
An ac modulation technique using a lock-in amplifier (SR830 Stanford Research Systems) was employed to record the differential conductance($\frac{dI}{dV}$) vs $V$ spectra. In this technique, a dc current (Keithley 6220) coupled with a small ac current (SR830 Stanford Research Systems) is passed through the point contact. The dc output voltage across the point contact is recorded using a  digital multimeter (Keithley 2000) and the ac output is measured by the SR830 lock-in amplifier. The first harmonic response of the lock-in signal will be proportional to $\frac{dV}{dI}$, which is used to calculate $\frac{dI}{dV}$. Data acquisition and analysis was done using LabVIEW based graphical user interface programmes.
\begin{figure}[h!]
 \includegraphics[scale = 0.5]{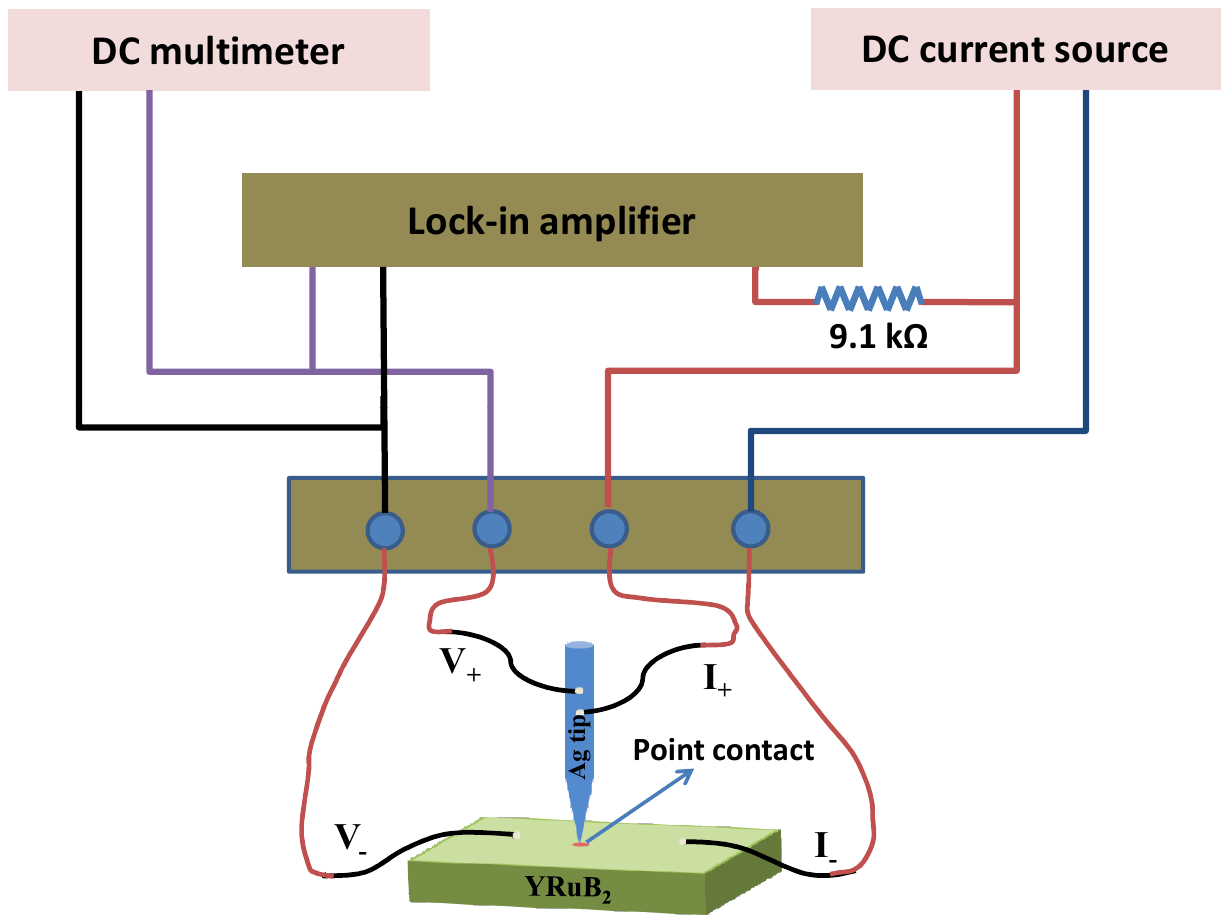}
 \caption{ Schematic describing the point contact spectroscopy measurements using lock-in based modulation technique.
  }
\end{figure}
\section{Analysis of Point contact Andreev reflection (PCAR) spectra:}
PCAR spectra are analyzed by the theory proposed by Blonder, Tinkham and Klapwijk (BTK) \cite{BTK}. According to BTK theory when the point contact is between a superconductor(S) and a normal metal(N), the current through the N/S point contact is given by\par

 I = C \( \int_{-\infty}^{+\infty} [f (E-eV) - f (E)][1+A(E)-B(E)] \,dE \)\par
 
where $f(E)$ is the fermi distribution function, $A(E)$ is the probability of Andreev reflection and $B(E)$ is the probability of normal reflection. From the above equation, PCAR spectra ($\frac{dI}{dV}$ vs $V$) can be simulated. The following formulas were used to calculate $A(E)$ and $B(E)$-
\\ 
For $E< \Delta$:\par
    $ A(E) = \frac{(\Delta/E)^2}{1-\varepsilon(1+2Z^2)^2} $\par
    and $B(E) = 1-A(E)$
\\
For $E>\Delta$:\par
    $ A(E) = \frac{u^2v^2}{\gamma^2} $\par
    and $B(E) = \frac{[u^2-v^2]^2Z^2(1+Z^2)}{\gamma^2}$\par
    where $u^2= 1-v^2=\frac{1}{2}[1+\frac{((E)^2-\Delta^2)^{1/2}}{E}]$, $\gamma^2=([u^2-v^2]Z^2+u^2)^2$, and $\varepsilon= \frac{(E^2-\Delta^2)}{E^2}$
\\
In order to incorporate the finite lifetime of the quasiparticle, BTK theory was modified by including a broadening parameter $\Gamma$\cite{Gamma}. The expressions for the coefficients $A(E)$ and $B(E)$ will be modified by replacing $E$ with $E+i\Gamma$. In modified BTK theory, $\Gamma$ along with the parameter Z and superconducting gap ($\Delta$) are the fitting parameters.\par
In the case of two gap superconductors, modified BTK theory is generalized to include two superconducting gaps\cite{twogap}. In this model, the current $I$ and hence the conductance ($\frac{dI}{dV}$) will be the sum of the conductance contribution from two different bands. The normalized conductance will be written as\par
$(\frac{dI}{dV})_N = w (\frac{dI}{dV})_{1N} + (1-w) (\frac{dI}{dV})_{2N}$

There are seven fitting parameters: superconducting gaps ($\Delta_1$ and $\Delta_2$), barrier strength ($Z_1$ and $Z_2$), broadening parameter ($\Gamma_1$ and $\Gamma_2$) and weight factor $w$. In this paper, all the conductance spectra are fitted within two gap BTK model with these seven fitting parameters.

\section{Estimation of point contact diameter:}
Contact diameter was measured using the Wexler formula given by\par
 $R_{PC} = \frac{2h/e^2}{(ak_F)^2} + \Gamma(l/a)\frac{\rho(T)}{2a}$\par
 Where $a$ is the contact diameter, $l$ is the electronic mean free path and $\Gamma(l/a)$ is a factor close to unity.
 
 The formula for contact resistance contains two terms -first term is the Sharvin resistance and the second term is the Maxwell resistance. In the ballistic regime where the contact diameter is much less than the elastic mean free path, the contribution to the resistance comes only from the first term in the Wexler formula. 

\begin{table}[H] 
\centering
\includegraphics[scale=0.7]{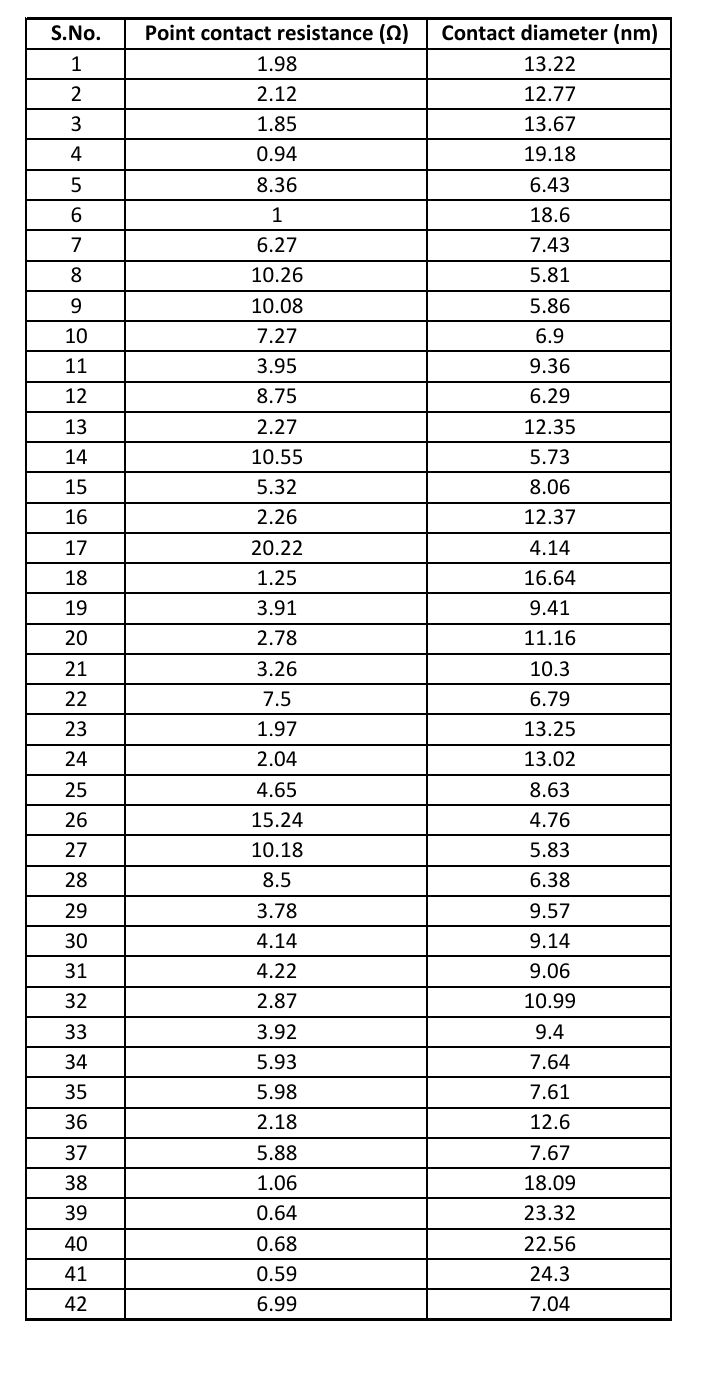}
\caption{A list of normal state resistance of the point contacts and respective estimated contact diameters.}
\end{table}
\section{Phonon dispersion and Eliashberg spectral function calculations:}
\begin{figure}[h!]
\centering
\includegraphics[width=0.95\textwidth]{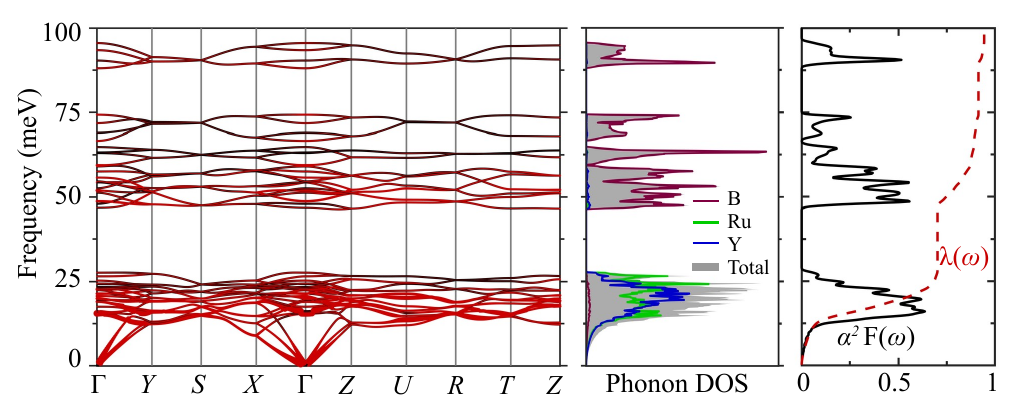}
\caption{The phonon dispersion along the various high-symmetric line (left), the total and partial phonon DOS (middle) and the Eliashberg spectral function ($\alpha^2F(\omega)$), and accumulated electron-phonon coupling constant $\lambda(\omega)$ (right). The size of the red circle in the phonon spectrum represents the strength of the electron-phonon coupling $\lambda_{{\bf q}\nu}$.}
\label{supple}
\end{figure}

The phonon dispersion, total and partial phonon DOS,  and Eliashberg spectral function $\alpha^2F(\omega)$ and cumulative contribution to the electron-phonon coupling strength $\lambda(\omega)$ are given in Figure~\ref{supple}. The sizes of the red dots on the phonon spectrum determine the strength of the electron-phonon coupling constant for each phonon band. Three acoustic branches near the $\Gamma$ point and a few low-lying optical modes have the dominating contribution to the electron-phonon coupling. The partial phonon DOS shows that the Y and Ru atoms contribute to the low-frequency phonon mode, whereas the high-frequency phonon modes are dominated by the lighter B atoms.
The integrated electron-phonon coupling constant $\lambda=2\int(\alpha^2F(\omega)/\omega)d\omega$ is 0.94. The superconducting critical temperature can be computed using the McMillan formula as modified by Allen and Dynes~\cite{Allen},
\begin{equation}
T_c=\frac{\omega_{\rm log}}{1.2}\exp\bigg[\frac{-1.04(1+\lambda)}{\lambda-\mu^*(1+0.62\lambda)}\bigg]
\label{eq1}
\end{equation}
where $\omega_{\rm log}=\exp[(2/\lambda)\int(\alpha^2F(\omega)/\omega){\rm ln}\omega d\omega]$ is the logarithmically averaged phonon frequency,  and $\mu^*$ is the Coulomb repulsion parameter whose value varies between 0.1 to 0.2 depending on the systems~\cite{margine2013anisotropic}.
The calculated value of $T_c$ is $8.7$ K using Eq.~\ref{eq1} with $\mu^* =0.2$.
\section{Scanning tunnelling microscopy/spectroscopy:}
\begin{figure}[H]
 \centering
 \includegraphics[width=0.63\textwidth]{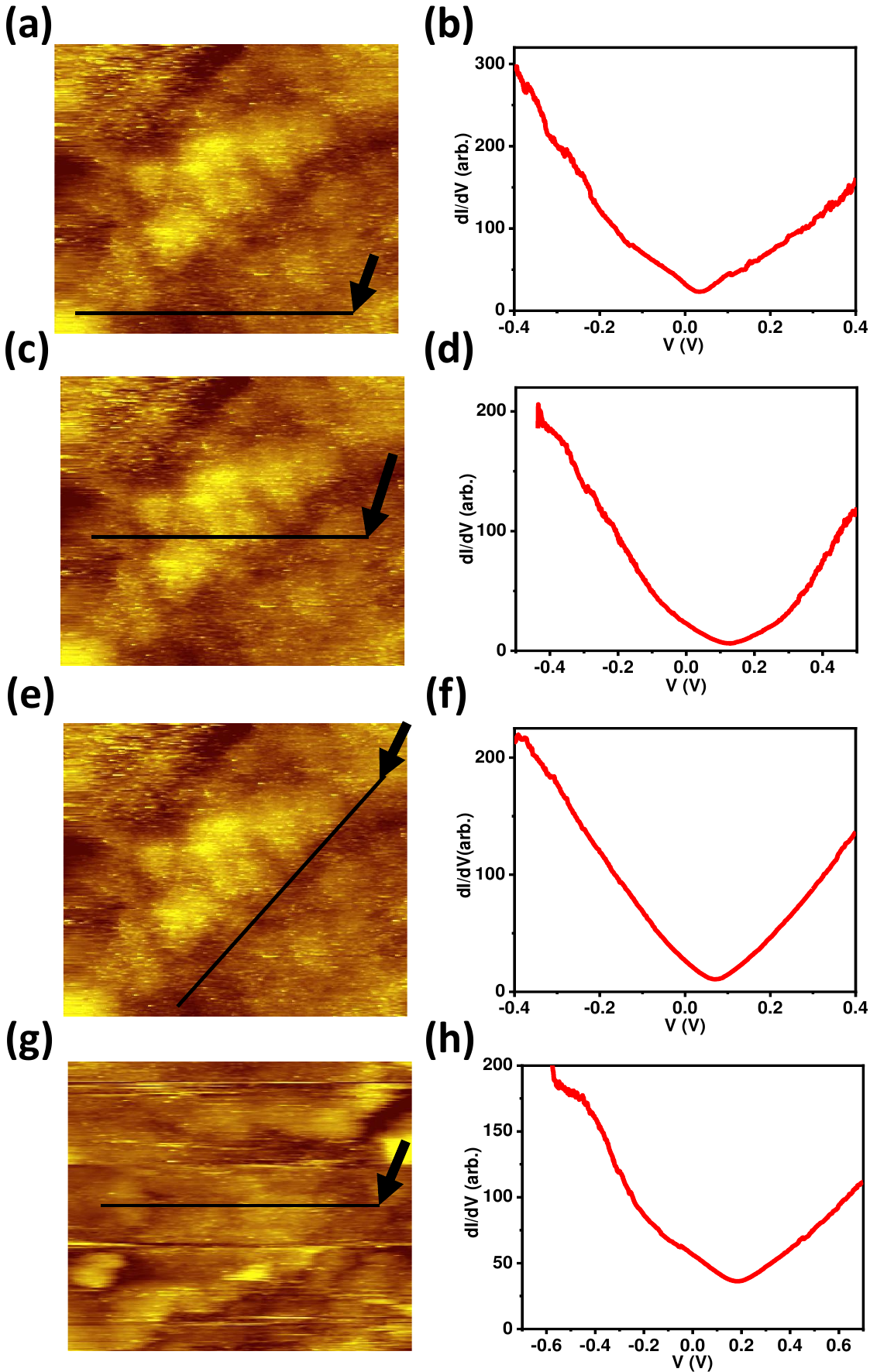}
 \caption{(a,c,e) STM images (20 $\times$ 20 nm) of reverse-sputtered YRuB$_2$ surface. (b,d,f) averaged scanning tunnelling spectroscopy (STS) spectra at 77 K obtained at 16 data points along a line as shown in (a, c and e) respectively (sample bias voltage = 0.65 V, tunneling current = 0.84 nA, lock in drive-amplitude = 65 mV). (g) STM image (30 $\times$ 30 nm). (h) Averaged STS spectra obtained at 16 data points along a line as shown in (g) (sample bias voltage = 250 mV, tunneling current = 0.85 nA, lock in drive amplitude = 25 mV). These STS spectra reveal presence of Dirac cone-like feature between 0 - 200 meV at different positions.}
\end{figure}
\pagebreak
\section{Scanning electron microscopy (SEM):}
\begin{figure}[H]
 \centering
 \includegraphics[width=0.98\textwidth]{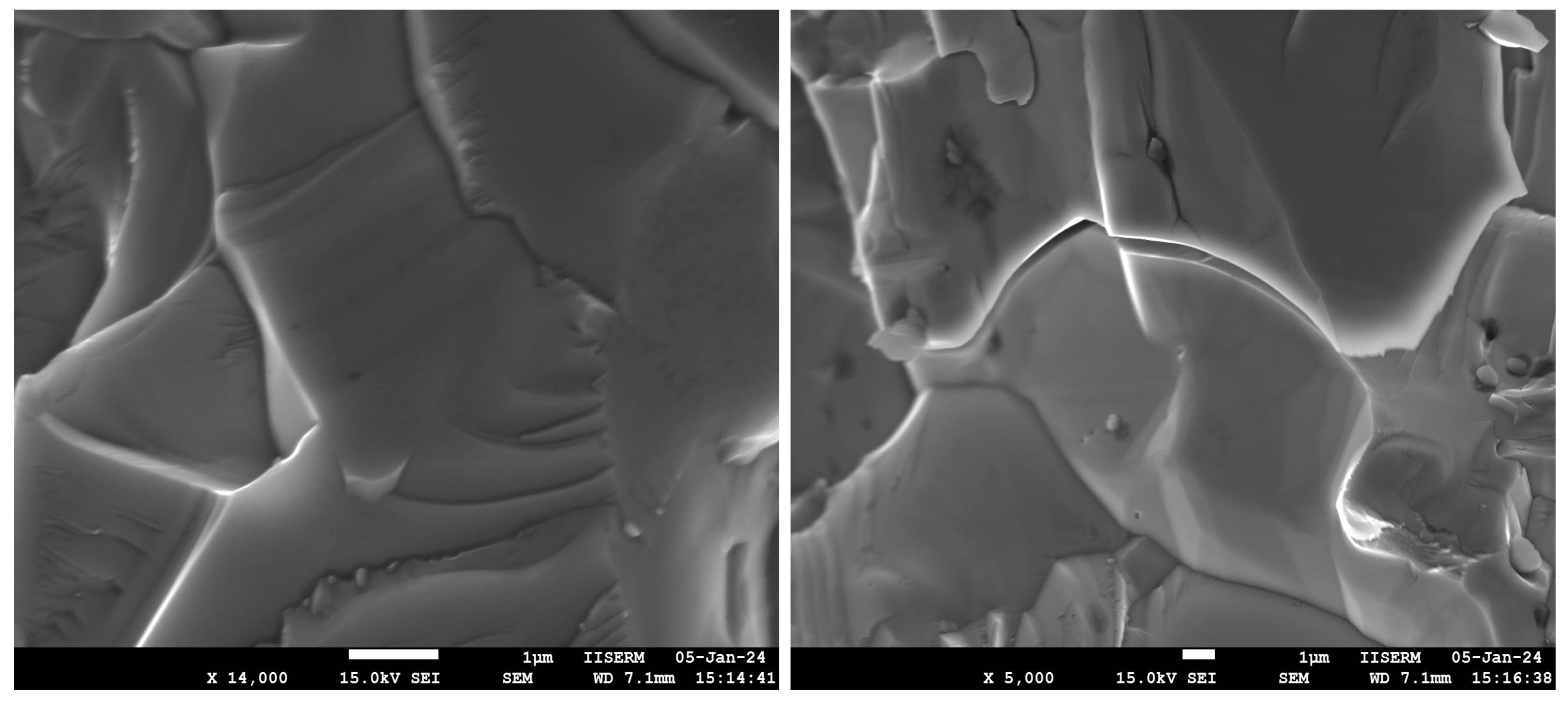}
 \caption{Scanning electron microscopy (SEM) images of polycrystalline YRuB$_2$.
  }
\end{figure}
\section{Resistivity:}
\begin{figure}[H]
 \centering
 \includegraphics[scale = 0.45]{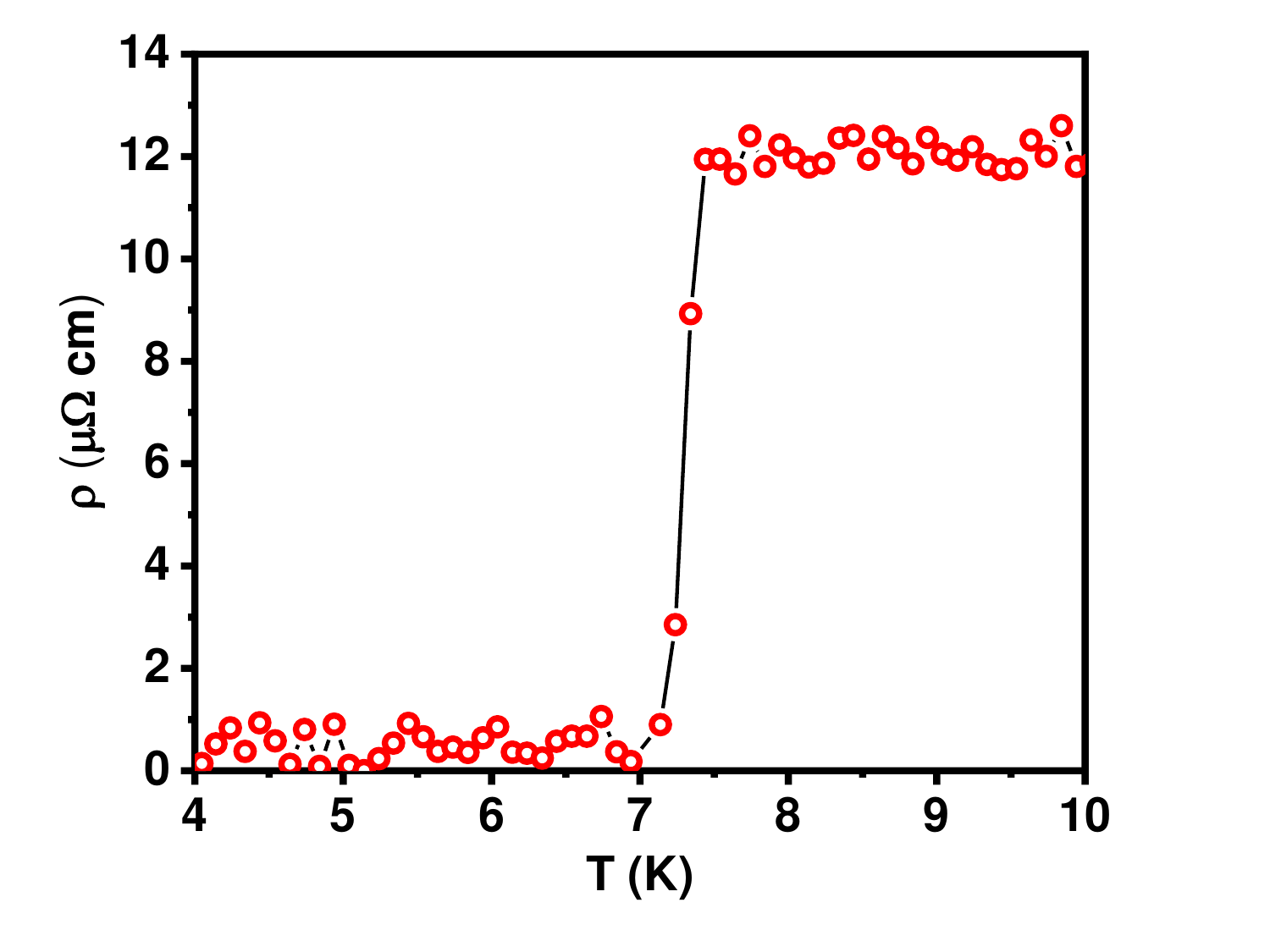}
 \caption{Temperature dependence of resistivity on polycrystalline YRuB$_2$ shows bulk superconductivity in YRuB$_2$.
  }
\end{figure}
\pagebreak
\section{Additional data:}
\subsection{Temperature and magnetic field dependence of PCAR spectra}

\begin{figure}[h!]
 \centering
 \includegraphics[scale = 0.65]{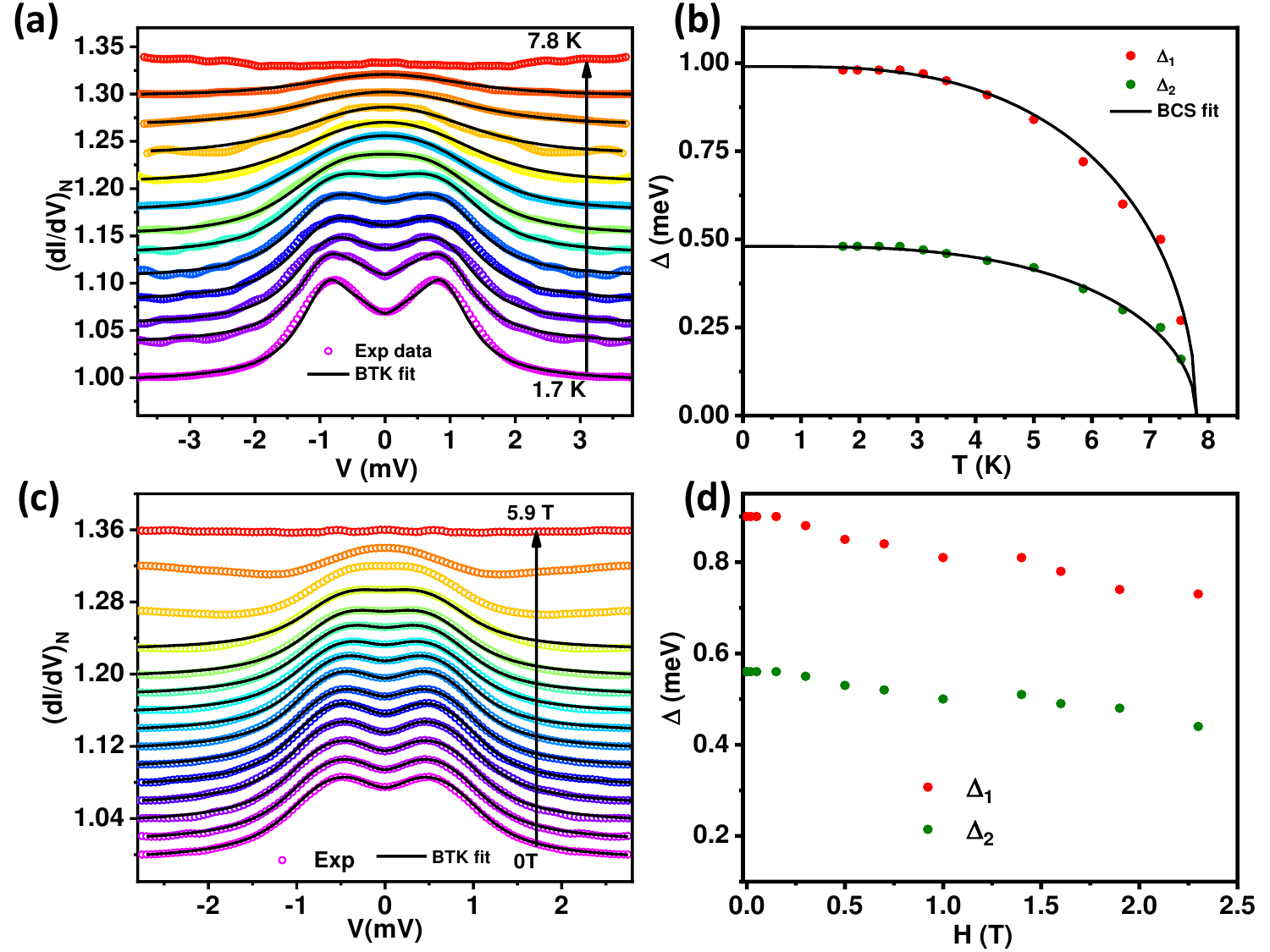}
 \caption{ (a) Temperature dependence of the conductance spectra. Temperature varies from 1.7 K up to the transition temperature. Colored circles represent the experimental data and the solid black line represents the theoretical fit using two-gap modified BTK theory (b) Variation of the extracted value of $\Delta_1$ and $\Delta_2$ with temperature. The black line shows the variation of $\Delta_1$ and $\Delta_2$ as per conventional BCS theory. Both the gaps follow conventional BCS-like behaviour. (c) Magnetic field dependence of the conductance spectra. The magnetic field varies from 0 T up to 5.9 T, where the conductance spectra became flat. Colored circles represent the experimental data and the solid black line represents the theoretical fit using two-gap modified BTK theory. (b) Magnetic field dependence of $\Delta_1$ and $\Delta_2$ extracted from the spectra shown in Figure 1(c).
  }
\end{figure}
\begin{figure}[H]
 \centering
 \includegraphics[width= 0.9\textwidth]{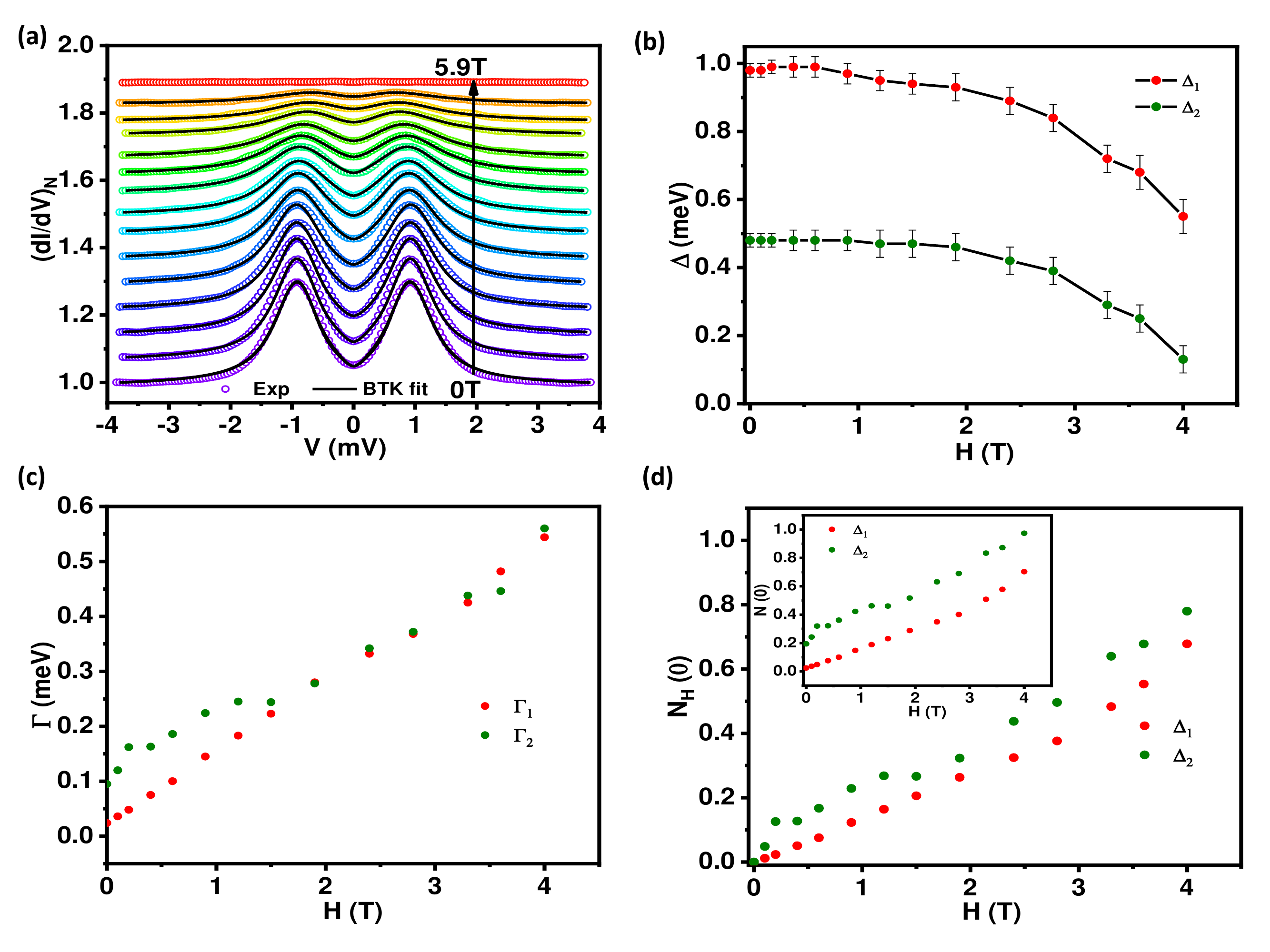}
 \caption{(a) Magnetic field ($H$) dependence of the conductance spectra (colored circles) with two gap BTK fit (black line) at 1.7 K. (b) Variation of the two superconducting gaps ($\Delta_1$ and $\Delta_2$ ) with $H$. (c) $H$ dependence of the broadening parameter $\Gamma_1$ and $\Gamma_2$. (d) $H$ dependence  of zero bias density of states ($N_H(0)$) after subtracting the zero field contribution corresponding to $\Delta_1$ and $\Delta_2$. The $inset$ shows the variation of zero energy density of state ($N(0)$) corresponding to $\Delta_1$ and $\Delta_2$.  
  }
  \label{fig_3}
\end{figure}
\pagebreak
\begin{figure}[htp]
 \centering
 \includegraphics[scale = 0.5]{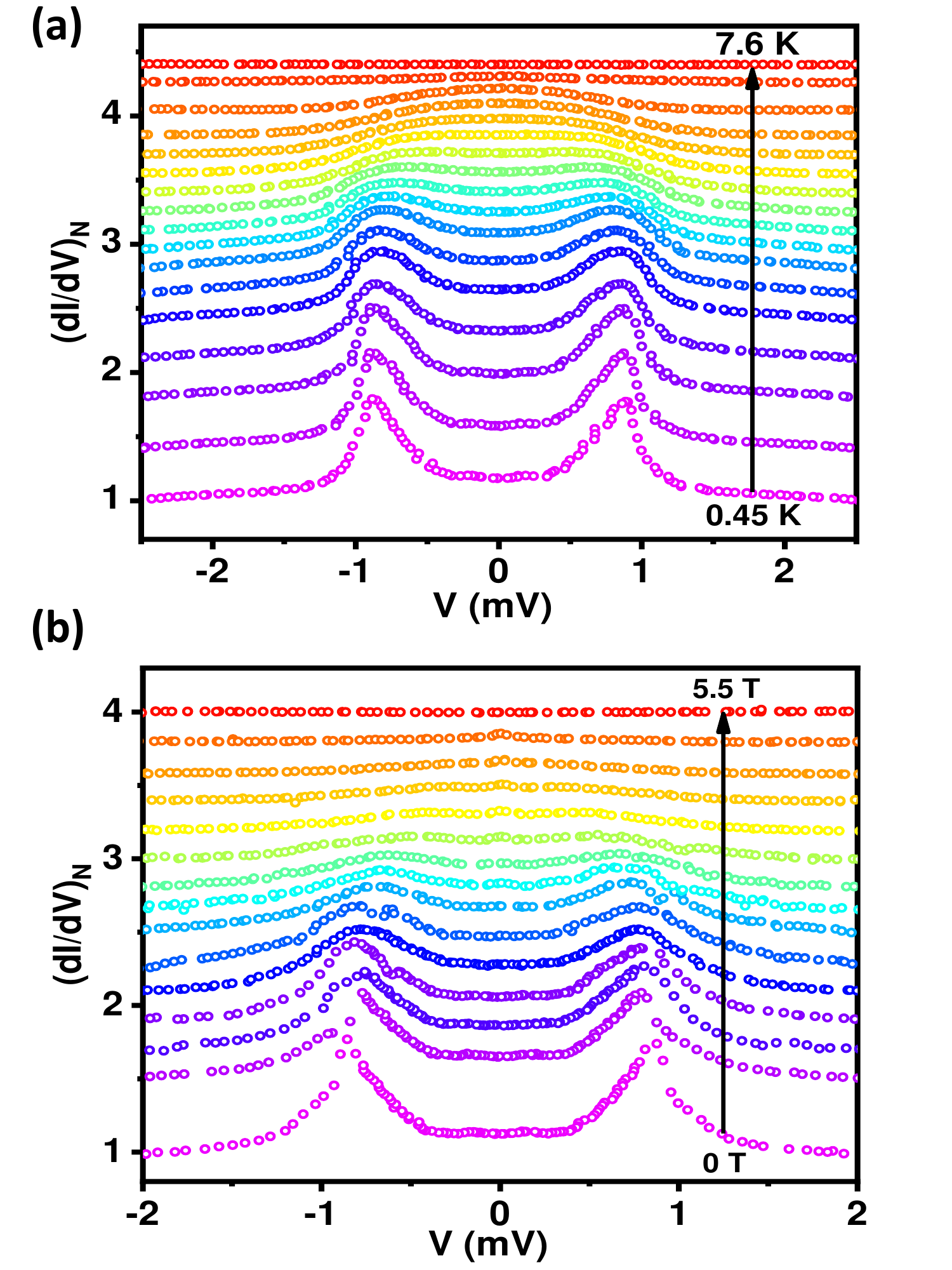}
 \caption{ (a) Temperature dependence of the conductance spectra. Temperature varies from 0.45 K up to the transition temperature. (c) Magnetic field dependence of the conductance spectra. The magnetic field varies from 0 T up to 5.5 T, where the conductance spectra became flat.}
\end{figure}
\pagebreak
\subsection{PCAR spectra}  
\begin{figure}[H]
 \centering
 \includegraphics[scale = 0.9]{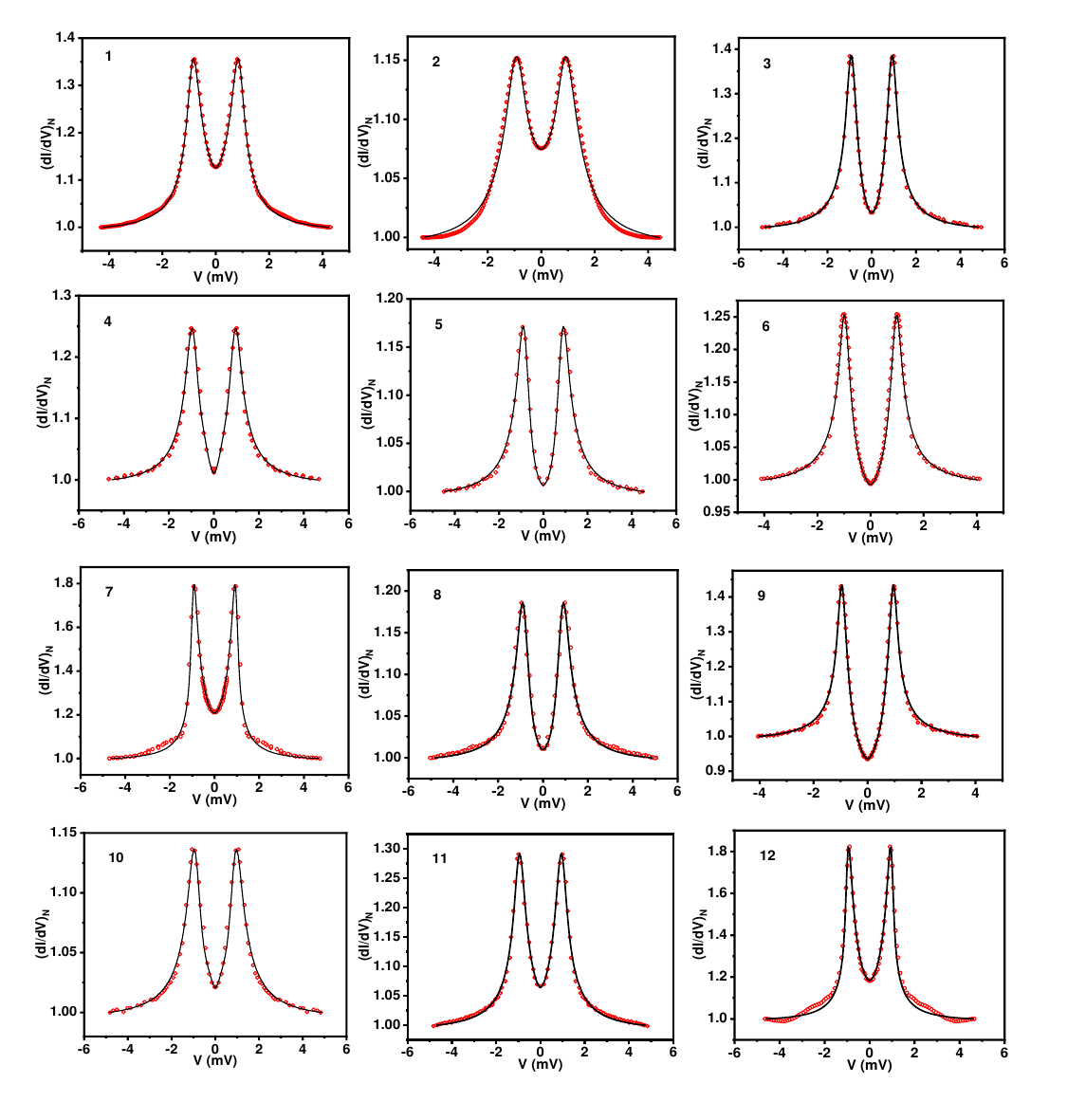}
 \caption{PCAR spectra recorded at different points on the surface of YRuB$_2$ using Ag tip, along with the two gap modified BTK fit (black line).}
\end{figure}
\begin{figure}[h!]
 \centering
 \includegraphics[scale = 0.9]{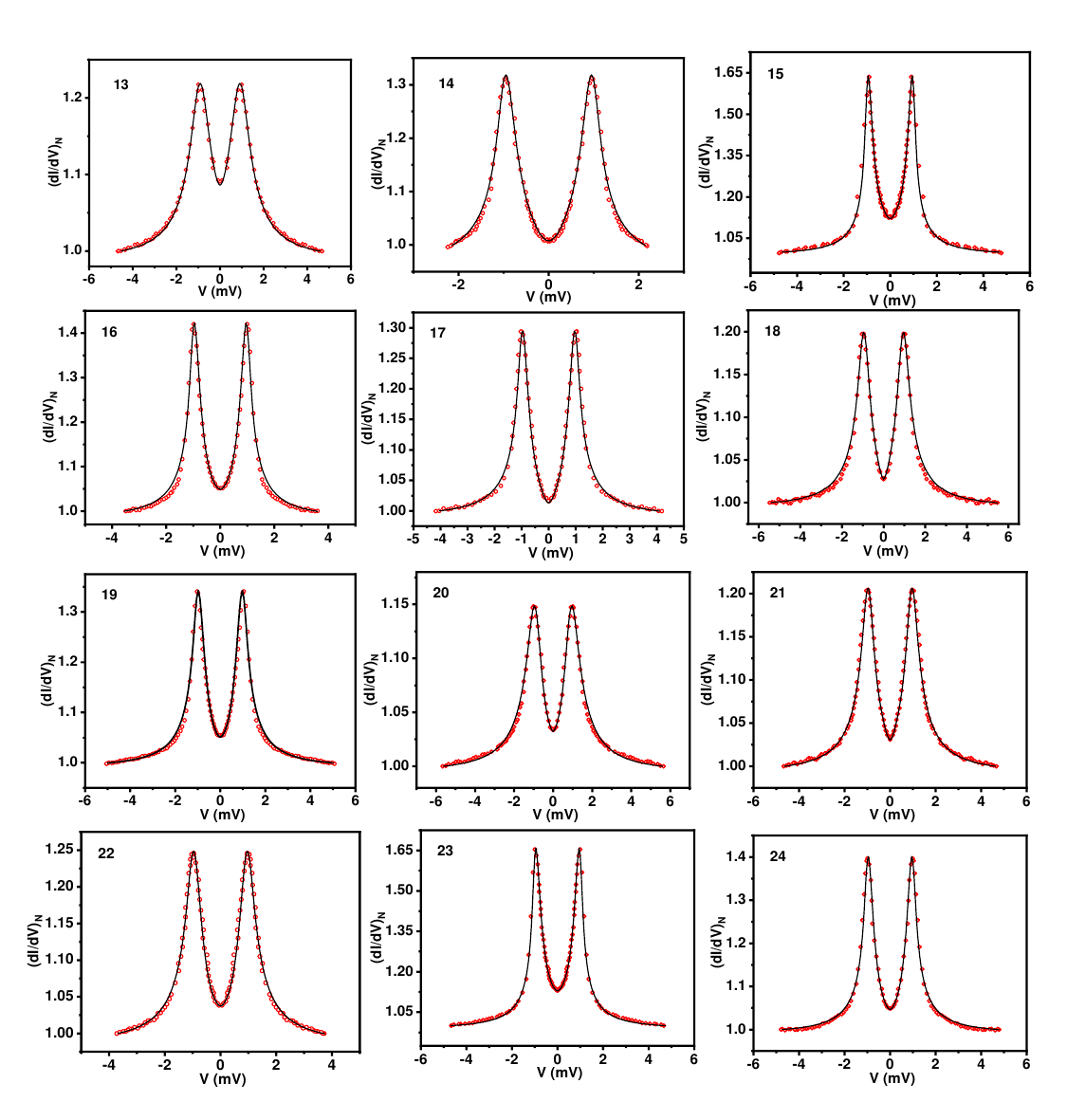}
 \caption{PCAR spectra recorded at different points on the surface of YRuB$_2$ using Ag tip, along with the two gap modified BTK fit (black line).}
\end{figure}
\begin{figure}[h!]
 \centering
 \includegraphics[scale = 0.9]{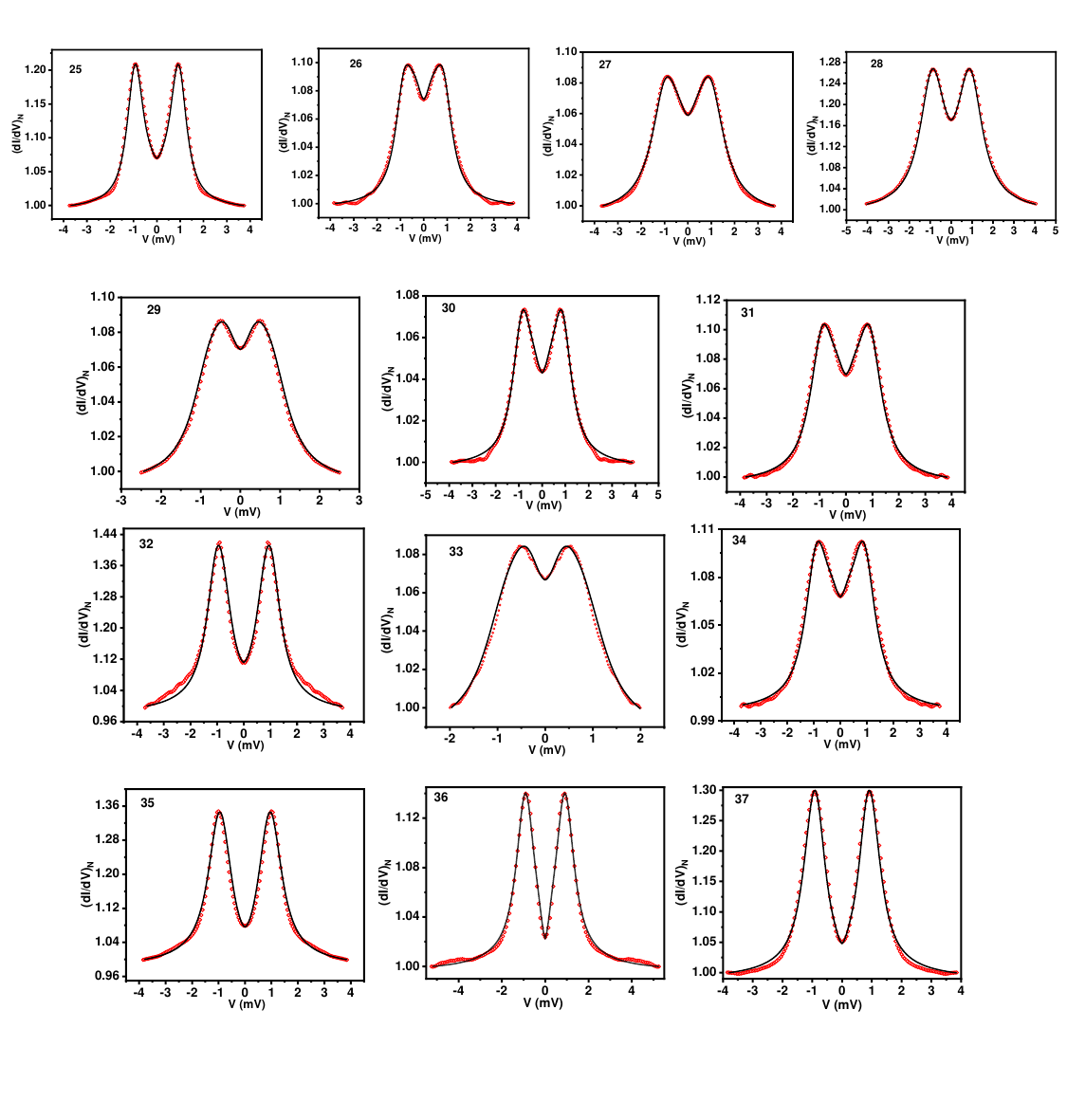}
 \caption{PCAR spectra recorded at different points on the surface of YRuB$_2$ using Ag tip, along with the two gap modified BTK fit (black line).}
\end{figure}
\pagebreak
\subsection{Fitting parameters:}
\begin{table}[H] 
\centering
\includegraphics[scale=0.68]{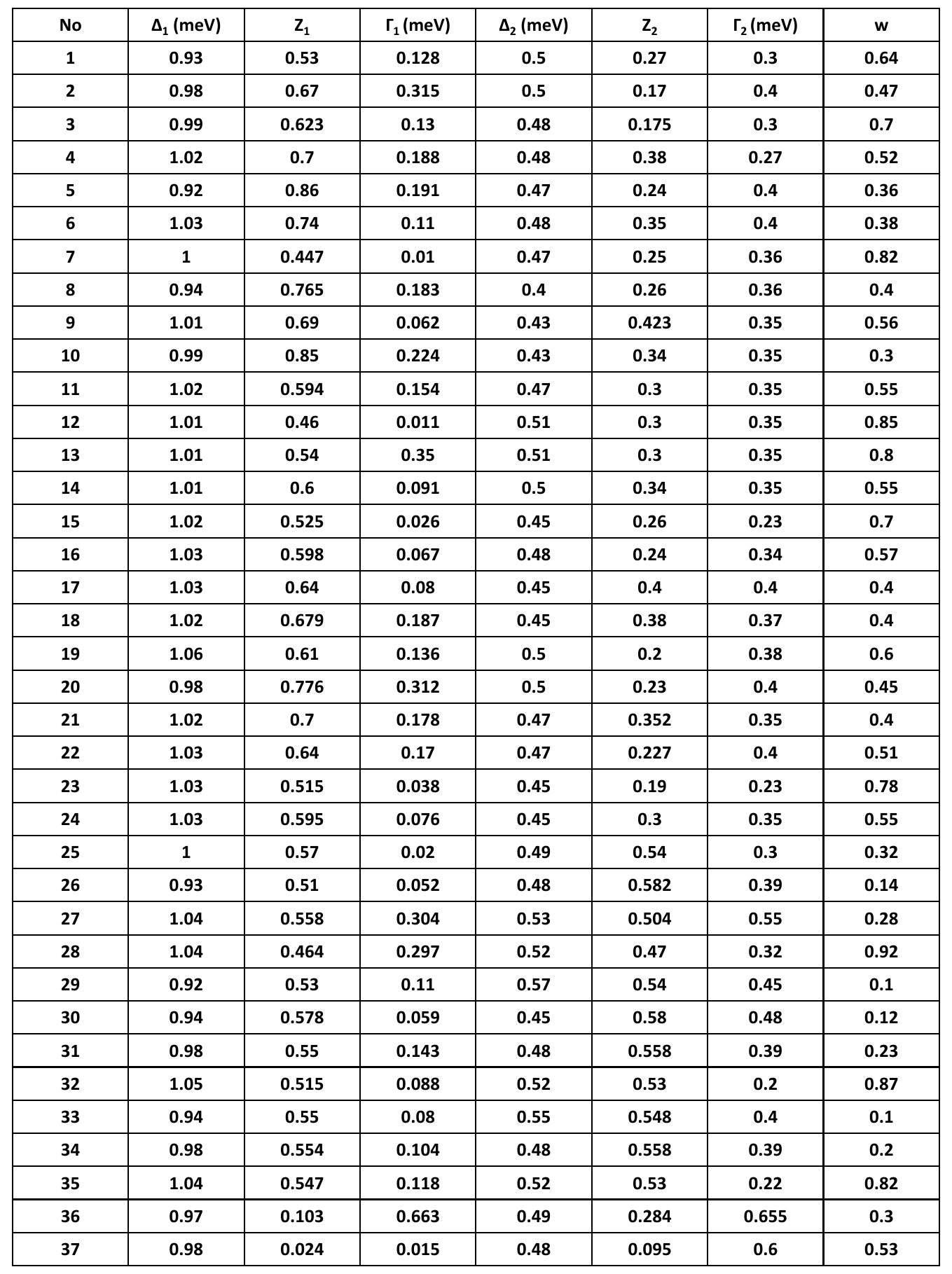}
\caption{Fitting parameter for the PCAR spectra shown in Figure S(6),S(7)\& S(8).}
\end{table}
\subsection{PCAR spectra showing two gap structure}  
\begin{figure}[h!]
 \centering
 \includegraphics[scale = 0.5]{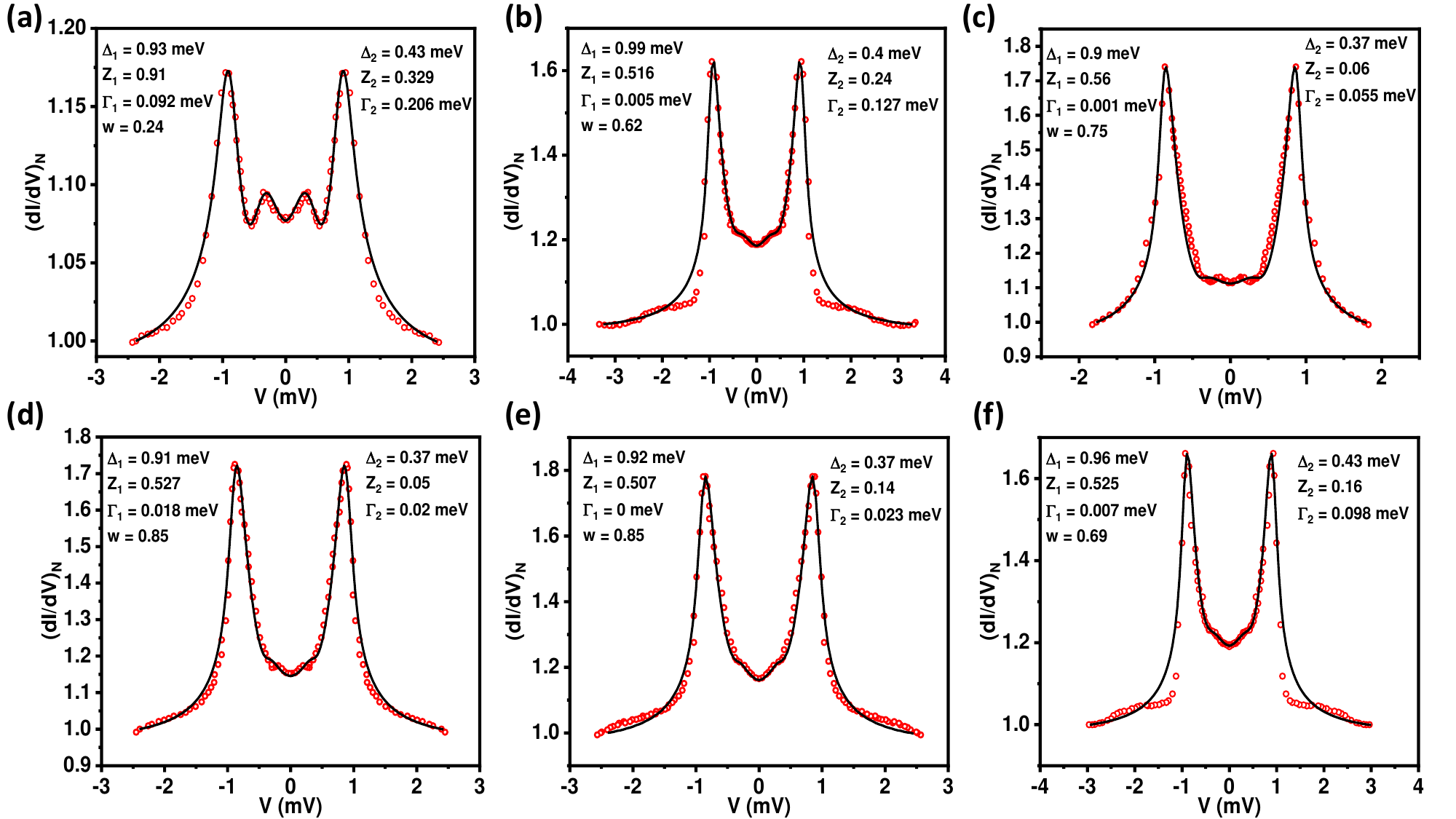}
 \caption{PCAR spectra recorded at different points on the surface of YRuB$_2$ using Ag tip, along with the two gap modified BTK fit (black line). PCAR spectra were recorded at $T$ $\sim$ 0.45 K.}
\end{figure}
\subsection{Zero bias density of states (N(0))}  
\begin{figure}[H]
 \centering
 \includegraphics[scale = 0.35]{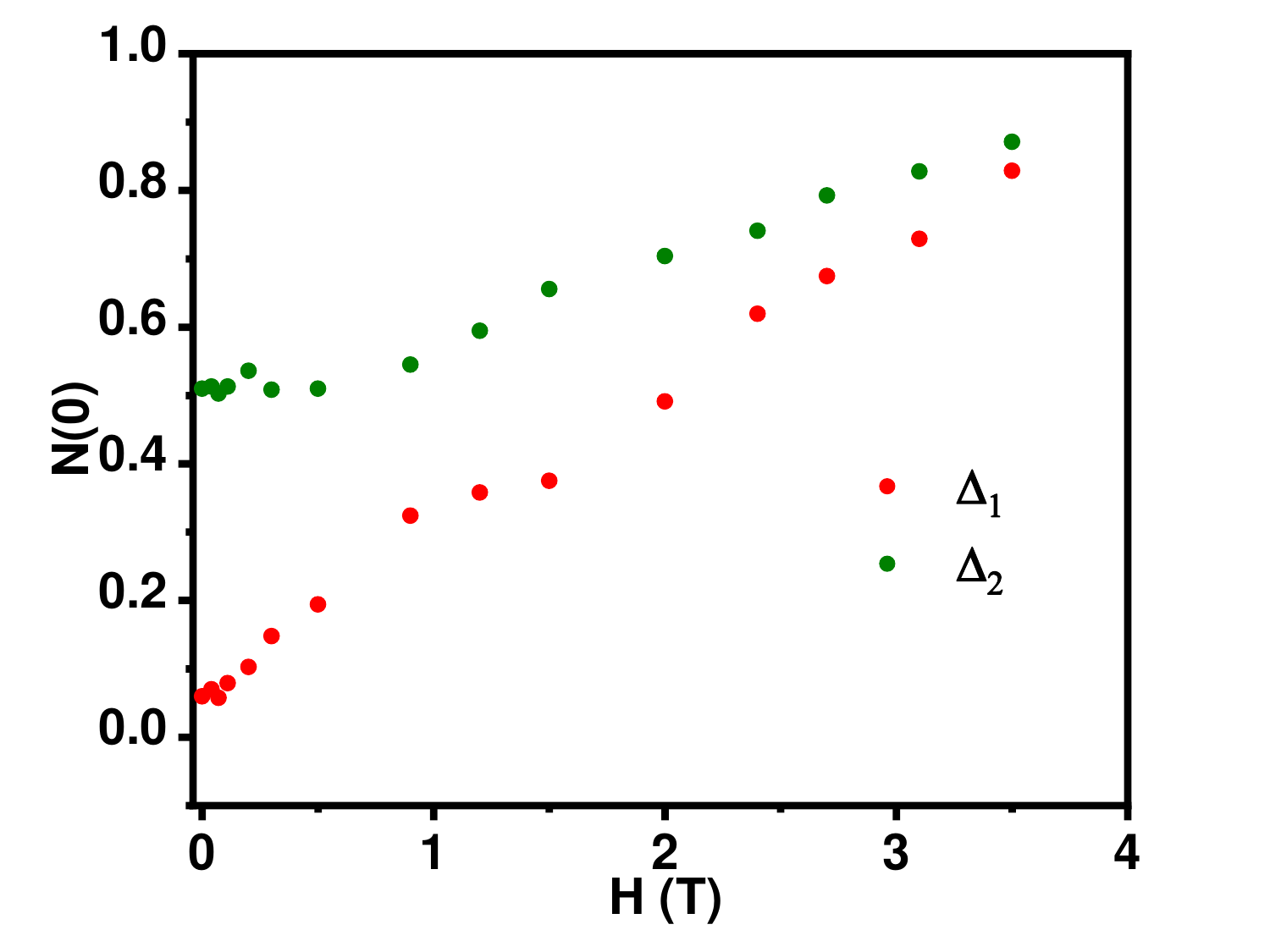}
 \caption{Magnetic field dependence of zero bias density of states (N(0))
corresponding to $\Delta_1$ and $\Delta_2$ for the spectra shown in
Figure 3(e) of main manuscript.}
\end{figure}


\end{document}